\documentclass[showpacs,letter,prl,a4,twocolumn]{revtex4}
\usepackage{epsfig}
\usepackage{subfigure}
\usepackage{amsmath}

\newcommand{\av}[1]{ {\left\langle #1 \right\rangle} }

\newcommand{\ket}[1]{ {| #1 \rangle} }
\newcommand{\bra}[1]{ {\langle #1 |} }
\newcommand{\vect}[1]{\vec{#1}}

\newcommand{\bs}[1]{\boldsymbol{#1}}

\begin{document}

\title[Fano-like Anti-resonances in Nanomechanical and Optomechanical Systems]{Fano-like Anti-resonances in Nanomechanical and Optomechanical Systems}%
\author{D.A. Rodrigues}
\affiliation{School of Physics and Astronomy, University of
Nottingham, Nottingham NG7 2RD, U.K.}%



\begin{abstract}
We study a resonator coupled to a generic detector and calculate the
noise spectra of the two sub-systems. We describe the coupled system
by a closed, linear, set of Langevin equations and derive a general
form for the finite frequency noise of both the resonator and the
detector. The resonator spectrum is the well-known thermal form with
an effective damping, frequency shift and diffusion term. In
contrast, the detector noise shows a rather striking Fano-like
resonance, i.e. there is a resonance at the renormalized frequency,
and an anti-resonance at the bare resonator frequency. As examples
of this effect, we calculate the spectrum of a normal state single
electron transistor coupled capacitively to a resonator and of a
cavity coupled parametrically
 to a resonator.
\end{abstract}

\maketitle



When a mechanical resonator is coupled to a detector, even weakly,
the detector can have a significant effect on the dynamics of the
resonator, and it is this ``back-action" that ultimately enforces
the standard quantum limit of measurement \cite{caves}. For weak
enough coupling, the detector acts like an additional thermal bath,
providing an effective frequency shift, damping and temperature.
This effective temperature can be lower than that of the resonator's
 environment, so a detector, such as an optical/microwave
cavity or a mesoscopic conductor, can cool the resonator
\cite{SSET-expt,cooling-expt}, potentially to its ground state
(\cite{MCGcoolreview} and refs. therein). Linear response theory
gives a general way of calculating the noise spectrum of a resonator
coupled to a detector, which is found to be very close to a thermal
spectrum \cite{Clerk}. A relevant question is if a thermal model is
enough to fully capture the dynamics of the system, or if there are
any effects beyond a purely thermal back action. In particular, when
calculating the spectrum of the detector, can we still treat the
back action as purely thermal?

 In this Letter, we
consider a detector linearly coupled to a resonator, and calculate
the noise spectrum of both. As expected, the resonator spectrum is
essentially thermal, with the frequency, damping and temperature
modified by the back-action. We might therefore expect that the
detector spectrum is close to the spectrum of a back-action-free
detector coupled to a resonator with this modified thermal bath.
 However, we find that this is not the case,
and show that the noise in the detector instead has a rather
striking feature akin to a Fano resonance \cite{fano}, {i.e.} a
resonance at the renormalized resonator frequency plus an
\emph{anti-resonance} at the original frequency. Fano resonances
arise from the interference between coherent and incoherent paths in
mesoscopic conductors \cite{fanoCWB}, but resonance/anti-resonance
pairs are a general interference phenomenon, occurring in systems
from LC circuits to electromagnetically induced transparency (EIT)
\cite{EITreview,classicalEIT}.

We first outline our general formalism before deriving an expression
for the spectrum of a generic detector coupled to a resonator.
Finally, we illustrate the analysis with two examples: a single
electron transistor coupled capacitively to a resonator, and a
cavity coupled parametrically to a resonator.

We assume that system (resonator plus detector) can be described
near a stable steady state by a linear set of Langevin equations,
\begin{eqnarray}
\dot{\vect{\varrho}}=-{\mathbf A}\vect\varrho+\vect\xi,
\label{eq:fullLV}
\end{eqnarray}
 where the vector
$\vect\varrho=(x,v,\sigma_1,\sigma_2...)$ includes the resonator
position $x$ and velocity $v$, and the degrees of freedom of the
detector $\sigma_i$. The term $\vect{\varrho}$ can refer to either a
set of classical dynamical variables, or a set of Heisenberg
operators for the system.
 The matrix $\mathbf A$ describes the evolution
of the means of the variables, and $\vect\xi$ describes fluctuations
about these means.
 If the equations of motion are non-linear, we can linearize
them to obtain the desired form, as long as the fluctuations of the
variables about their steady state values are small \cite{WM}.

 The
noise in the system can be calculated from Eq. (\ref{eq:fullLV})
directly, but by considering the sub-systems separately and assuming
a specific form for the coupling we find that we can write down a
more illuminating general form for both resonator and detector
noise. The resonator (either mechanical or a superconducting
stripline or coplanar waveguide \cite{cooling-expt}) is assumed to
be a single-mode harmonic oscillator, whose motion depends on a
linear sum of the detector variables, which we denote $\sigma_n$,
with overall strength $x_s$. The resonator variables then obey the
equations of motion $\dot x=v$ and,
\begin{eqnarray}
\dot{v}&=&-\omega_0^2x -\gamma_{e}v+\omega_0^2 x_s \sigma_n
+\xi_{e}, \label{eq:eomxv}
\end{eqnarray}
with $\omega_0$ the resonator frequency and $\gamma_{e},\xi_e$
 the damping and fluctuations it feels due to its
thermal bath.

The detector is described by a set of linear equations for its
degrees of freedom, $\sigma_i$,
\begin{eqnarray}
\dot{\vect{\sigma}}&=&-\mathbf{B}\vect\sigma+\vect\kappa
\;x+\vect{\xi}_\sigma.
  \label{eq:eomsigma}
\end{eqnarray}
Matrix $\mathbf{B}$ gives the $x$-independent evolution of the
detector variable means. The rate of change of each variable
$\sigma_i$ depends on resonator position $x$ with strength
$\kappa_i$.

 The vector
$\vect{\xi}_\sigma$ in Eq. (\ref{eq:eomsigma}) describes the
fluctuations of $\vect \sigma$, which may simply be thermal but can
also arise from other stochastic processes (such as incoherent
tunneling  \cite{RM}). In the Markovian approximation, the $\xi$
terms acting on both the resonator and the detector will be
$\delta$-correlated, $\av{\xi_i}=0$,
$\av{\xi_i(t)\xi_j(t')}=G_{ij}\delta(t-t')$. The correlators are
related to the variances  by \cite{WM},
\begin{eqnarray}
\mathbf{G}=(\mathbf{A}\bs\chi+\bs\chi \mathbf{A^T})\label{eq:defG},
\end{eqnarray}
with $\bs\chi$ the matrix of steady state variances of the coupled
system
$\chi_{ij}=\av{\varrho_i\varrho_j}-\av{\varrho_i}\av{\varrho_j}$. In
a quantum system, the commutation relations can mean that
$\chi_{ij}\neq\chi_{ji}$.

%

With our formalism set up, we solve the equations of motion. Fourier
transforming Eq. (\ref{eq:eomsigma}) and solving for $\sigma_n$,
\begin{eqnarray}
\sigma_n(\omega)=\vect{n}^T(\mathbf{B}+i\omega)^{-1}\vect\xi_\sigma(\omega)+x(\omega)\vect{n}^T(\mathbf{B}+i\omega)^{-1}\vect\kappa,
\label{eq:solFTsigma}
\end{eqnarray}
%
where $\vect{n}^T$ is the row vector defined by
$\vect{n}^T\vect\sigma=\sigma_n$. We substitute this into the
Fourier transform of Eq. (\ref{eq:eomxv}) to get $x(\omega)$. We can
then obtain the resonator spectrum, using Eq. (\ref{eq:defG}), the
Hermitian nature of the fluctuations, and the fact that noise
operators $\delta$-correlated in time will also be
$\delta$-correlated in frequency
$\av{\xi(\omega)\xi(\omega')}\propto\delta(\omega+\omega')$.
Neglecting resonator bath fluctuations $\xi_e$,
\begin{eqnarray}
S_{xx}(\omega)=\frac{S_{vv}(\omega)}{\omega^2}&=&
\frac{\omega_0^4x_s^2S_{\sigma_n\sigma_n}'(\omega)}
{\left(\omega_r(\omega)^2-\omega^2\right)^2+\omega^2\gamma_{T}(\omega)^2}.\label{eq:Svv}
\end{eqnarray}
This is written in the form of a resonator coupled to a heat bath,
 defining the
renormalized frequency $\omega_r^2(\omega)=
\omega_0^2+\delta\omega_0^2(\omega)$ and total damping
$\gamma_{T}(\omega)=\gamma_{e}+\gamma_\sigma(\omega)$. Noting that
the frequency-dependent effective back-action damping
$\gamma_\sigma(\omega)$ and frequency shift
$\delta\omega_0^2(\omega)$ terms are real, these expressions are
given by \cite{foot_1},
\begin{eqnarray}
\delta\omega_0^2(\omega)+i\omega\gamma_\sigma(\omega)&=&-
\omega_0^2x_s\vect{n}^T(\mathbf{B}+i\omega)^{-1}\vect\kappa
\label{eq:gplusw}.
\end{eqnarray}
    $S_{\sigma_n\sigma_n}'(\omega)$ is closely related to
$S_{\sigma_n\sigma_n}^0(\omega)$, the zero-coupling spectrum of
$\sigma_n$, and is given by,
\begin{eqnarray}
\mathbf{S}_{\sigma\sigma}'(\omega)&=&(\mathbf{B}+i\omega)^{-1}\mathbf{G}_{\sigma\sigma}(\mathbf{{B}^T}-i\omega)^{-1}\label{eq:modS0}
\end{eqnarray}
where $\mathbf{G}_{\sigma\sigma}$ refers to the subset of
$\mathbf{G}$ that describes the detector variables. The resonator
can modify $\mathbf{G}_{\sigma\sigma}$ and linearization can
renormalize $\mathbf B$, but for weak enough coupling both these
effects will be small. If we can approximate these terms by their
uncoupled values, then we have $S_{\sigma_n\sigma_n}'(\omega)\approx
S_{\sigma_n\sigma_n}^0(\omega)$, and the noise on the resonator
reduces to the standard linear response expression. Equation
(\ref{eq:Svv}) describes a thermal-like spectrum, where weak
coupling implies a narrow resonance, so we can approximate the
$\omega$-dependent terms by their values at
$\omega_r\approx\omega_0$,  and define a diffusion term
$D_{\sigma}=\omega_0^4x_s^2S_{\sigma_n\sigma_n}'(\omega_0)$.

%

The effect of the detector back-action on the resonator spectrum is
essentially just a modification of the thermal parameters. We might
therefore expect the detector spectrum to be basically that of a
back-action-free detector coupled to a resonator with modified
thermal parameters, $\omega_r, \gamma_{T},D_\sigma$. However, we
find that that this is not the case, and that the detector noise can
be strongly modified from this naive picture,
\begin{equation}
S_{\sigma_n\sigma_n}(\omega)=
\frac{\left(\omega_0^2-\omega^2\right)^2+\omega^2\gamma_{e}^2}
{\left(\omega_r(\omega)^2-\omega^2\right)^2+\omega^2\gamma_{T}(\omega)^2}S_{\sigma_n\sigma_n}'(\omega).\label{eq:Snn}
\end{equation}
The spectrum of the detector noise has a Fano-like resonance
\cite{fano,foot_2}, i.e. there is a resonance at the renormalised
frequency $\omega=\pm\omega_r(\omega)$ and an \emph{anti-resonance}
at the \emph{un}renormalised frequency $\omega=\pm\omega_0$.
Somewhat surprisingly, we find that if $\gamma_{e}=0$ the noise at
the resonator frequency is exactly zero, {i.e.}
$S_{\sigma_n\sigma_n}(\pm\omega_0)=0$, independent of the parameters
of the detector. Just as the spectrum of the resonator near
$\omega_0$ is captured by three parameters,
$\omega_r,\gamma_T,D_\sigma$, the spectrum of $\sigma_n$ only
requires the additional two parameters $\omega_0, \gamma_{e}$.

The anti-resonance in the detector noise can be understood in a
simple, intuitive way \cite{classicalEIT}. The detector noise at
frequency $\omega$ depends on how sensitive the detector is to an
external oscillating force at that frequency.
For our coupled system, a perturbation on the detector at frequency
$\omega$ will also cause the resonator to respond at $\omega$. The
detector then feels the perturbation in two ways: the original
force, and a corresponding force from the resonator. The resonator
force is exactly out of phase with the external perturbation and so
acts to cancel it. When $\gamma_e=0$, the forces exactly cancel and
the noise goes to zero at $\omega=\omega_0$. Thus the anti-resonance
can be understood as classical interference, analogous to quantum
interference in mesoscopic conductors \cite{fanoCWB} or EIT
\cite{classicalEIT}.

We also see why the sub-systems have such a different effect on each
other.
The detector relaxes quickly (compared to the resonator damping), so
has a short ``memory" and acts like a Markovian thermal bath for the
resonator. In contrast, weak damping $\gamma_e$ means the resonator
has a long memory, so the detector sees a highly non-Markovian bath
and has a non-thermal spectrum.

Although no  approximations have been made in deriving Eqs.
(\ref{eq:Svv}) and (\ref{eq:Snn}) from Eq. (\ref{eq:fullLV}),
$\gamma_\sigma,\delta\omega_0^2$ and $S_{\sigma_n\sigma_n}'$ are all
$\omega$-dependent, so generally Eqs. (\ref{eq:Svv}) and
(\ref{eq:Snn}) will not look like thermal and Fano spectra. These
emerge only when the coupling between the resonator and detector is
weak enough that the resonance is narrow on scale of the change in
these parameters, i.e. $\gamma_T(\omega_0)$ must be much smaller
than the dissipative terms on the detector.

We now include thermal fluctuations from the resonator's
environment. Assuming that the resonator's bath is unaffected by the
detector, Eq. (\ref{eq:defG}) shows that
 the bath flucuations $\xi_e$ obey the standard expressions,
$\av{\xi_e}=0$, $\av{\xi_e(t)\xi_e(t')}=\delta(t-t')D_e$, where
$D_e$ is the diffusion.
We also require the correlations between the fluctuation terms on
the resonator $\xi_e$ and on the detector $\xi_i$. If these terms
vanish  $\av{\xi_e(t)\xi_i(t')}=0$, (typically the case
\cite{foot_3}), the resonator is described by Eq. (\ref{eq:Svv})
with a total diffusion $D_T=D_\sigma+D_e$. The detector spectrum is,
\begin{eqnarray}
S_{\sigma_n\sigma_n}(\omega)&=&
\frac{\left(\omega_0^2-\omega^2\right)^2+\omega^2\gamma_{e}^2 }
{\left(\omega_r(\omega)^2-\omega^2\right)^2+\omega^2\gamma_{T}(\omega)^2}S_{\sigma_n\sigma_n}'(\omega)\nonumber\\
&&+
\frac{\delta\omega_0^2(\omega)^2+\omega^2\gamma_{\sigma}(\omega)^2}
{\left(\omega_r(\omega)^2-\omega^2\right)^2+\omega^2\gamma_{T}(\omega)^2}\frac{D_e}{\omega_0^4x_s^2}.
\label{eq:S_Dn}
\end{eqnarray}
 The detector noise at
$\omega_0$ is suppressed below the uncoupled value by the factor
$R={S_{\sigma_n\sigma_n}(\omega_0)}/{S_{\sigma_n\sigma_n}^0(\omega_0)}\sim(\gamma_{e}^2+\gamma_\sigma^2\frac{D_e}{D_\sigma})/\gamma_T^2$.
Thus $\gamma_e^2/\gamma_\sigma^2\ll 1$ is a neccessary condition for
$R\ll1$.
  In terms of the effective thermal
occupation numbers, when the back action is strong enough that
$\frac{n_\sigma}{n_e}\gg\frac{\gamma_{\sigma}}{\gamma_{e}}\gg1$, Eq.
(\ref{eq:S_Dn}) reduces to Eq. (\ref{eq:Snn}) and $R\approx
\gamma_{e}^2/\gamma_\sigma^2\ll 1$. In the opposite regime $n_\sigma
\lesssim n_e$, $\gamma_e\ll \gamma_\sigma$, the suppression of the
noise is given by $R\approx\frac{ \gamma_e(2 n_e+1)}{
\gamma_\sigma(2n_\sigma+1)}$. So, the anti-resonance is only present
if the back action dominates, i.e. $n_\sigma\gg
n_e\gamma_\sigma/\gamma_e$, or the parameters allow ground state
cooling. Thus, the effect will be hard to observe when the the
resonator is cooled from a very high ambient temperature, as is
often the case with optical frequency cavities. Fig. 1(a) inset
($iii$) shows the effect of including the environment using
parameters from Ref. \cite{SSET-expt}, (where
$\gamma_e\approx\gamma_\sigma$, $T_\sigma \approx 100mK$ and
$T_e\lesssim 600mK$).

\begin{figure}
 {\vspace{-0.5cm}{\includegraphics[width=9cm]{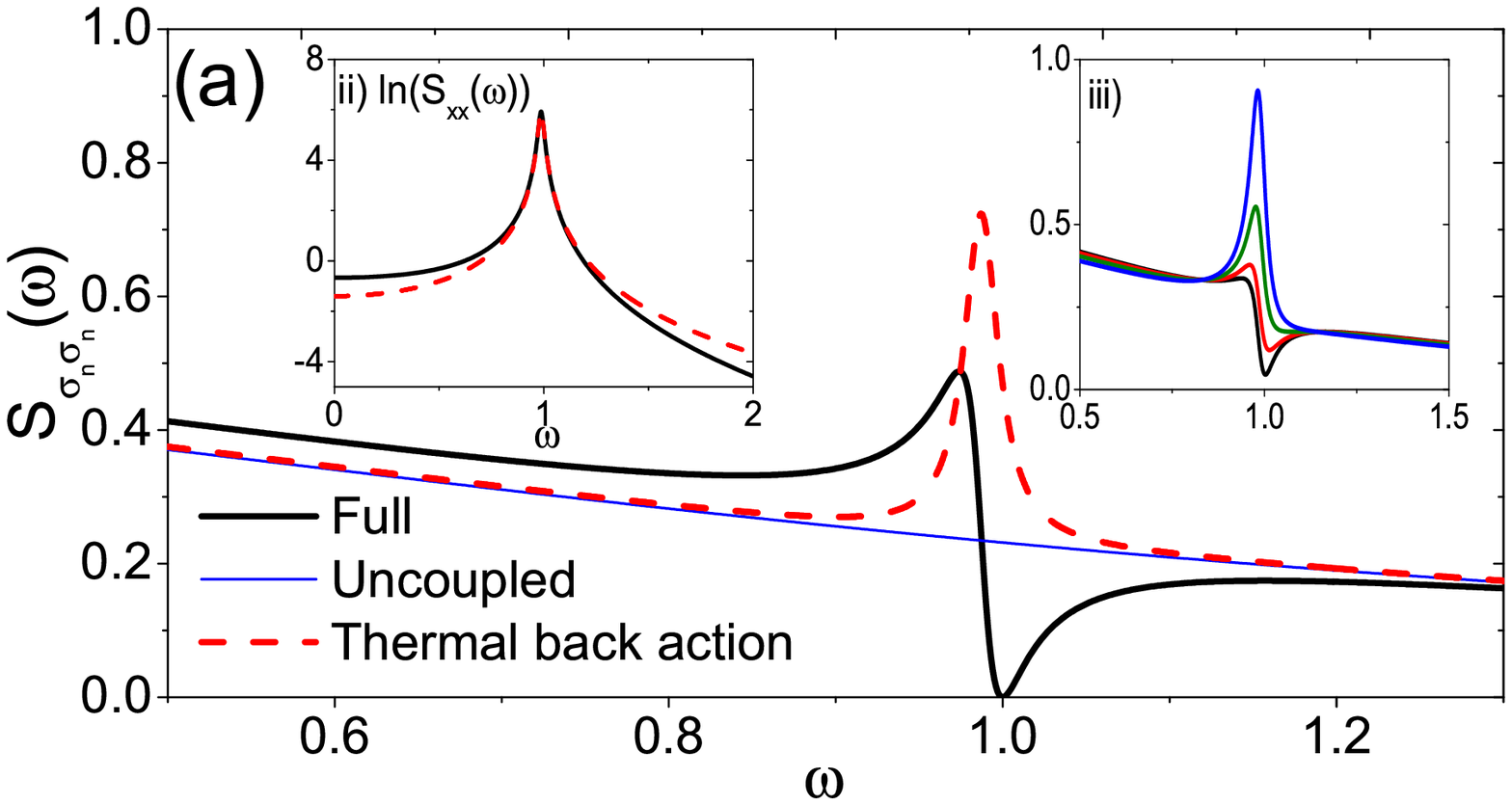}}
 \vspace{-1.5cm}}
{ {\includegraphics[width=9cm]{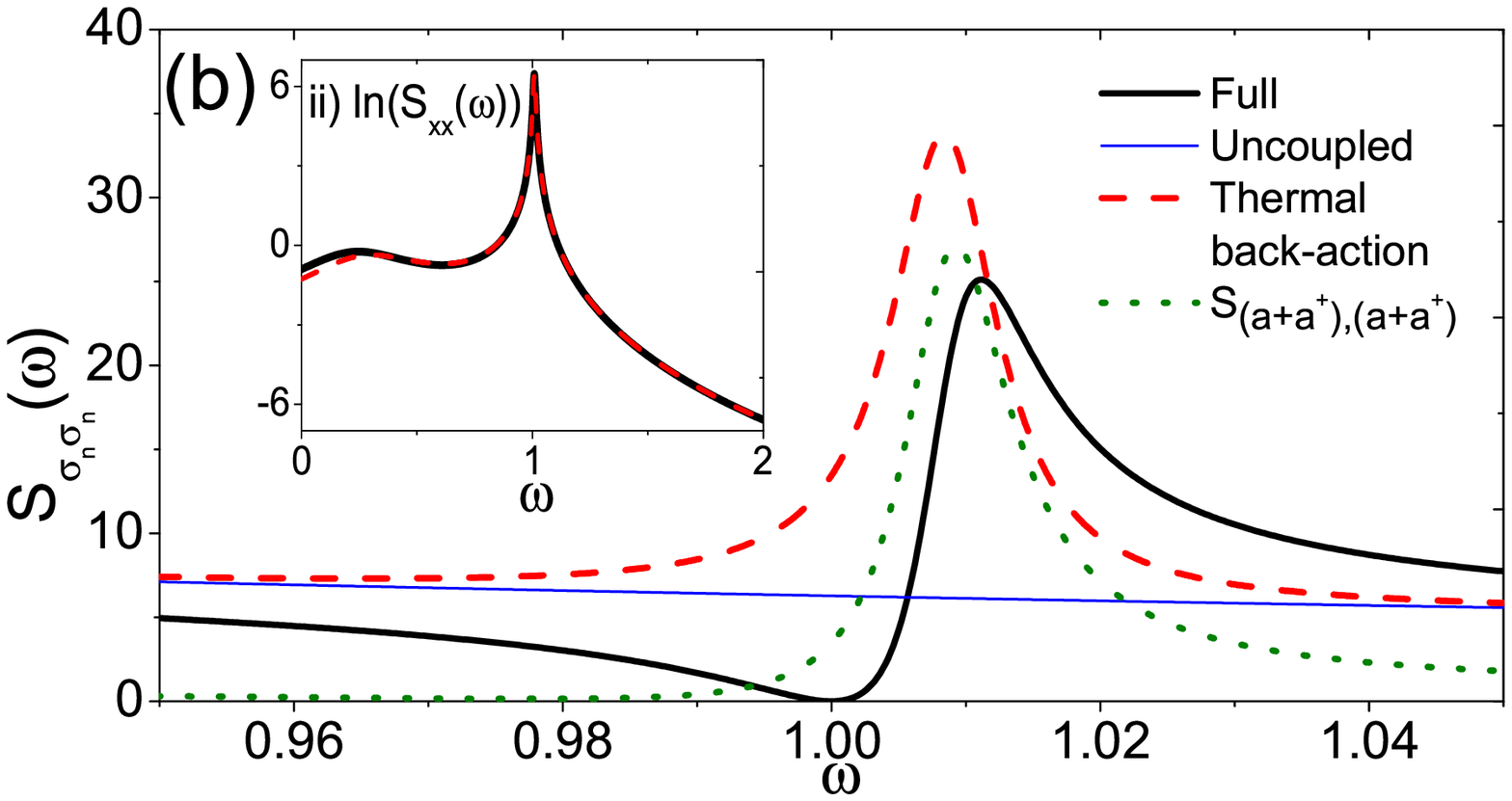}}\vspace{-0.5cm}}
 \caption{(color
online). Fano-like anti-resonances in the spectra of an SET(a) and a
cavity(b) coupled to a resonator. Shown is detector noise in the
fully coupled system (solid line), when the coupling is zero (thin
line) and when the back action is assumed to be purely thermal
(dashed line). The thermal model fails to capture the
anti-resonance, although it matches the resonator noise very well
(insets $ii$, logarithmic scale). The SET plot also shows the effect
of environmental diffusion and damping for experimentally relevant
values $(iii)$ \cite{SSET-expt}. The noise of a general quadrature
$a+a^\dag$ (rather than
$\sigma_n=a'\langle{a^\dag}\rangle+{a'}^\dag\langle{a}\rangle$)
shows a signature of the anti-resonance in its asymmetric peak
(dotted). SET parameters: $\Gamma_L=0.6,\Gamma_R=0.4, \omega_0=1,
\kappa=0.05$ and $x_s=1$; for the cavity: $\Delta=0.3, \gamma_a=0.5,
g=0.05, \omega_0=1, N_a=0$ and $\alpha=1$; for both main plots
$\gamma_e$ and $D_e=0$ and for inset $(iii)$  (a),
$\gamma_e/\gamma_\sigma=1$ and $D_e/D_\sigma=6,3,1$ and $0$ from
 top to bottom  \cite{SSET-expt}. } \label{fig:spectra}
\end{figure}


We now illustrate this formalism by two well-known systems, a
resonator coupled to a normal-state SET, and to an optical or
microwave cavity. In the SET case \cite{SETa,SSET1}, the resonator
position couples to a single operator describing the charge on the
island, $\sigma=\ket{1}\bra{1}$, where $\ket{1}$ is the state with
one extra charge on the island. If the junction resistances are
equal, we have a single, closed, linear equation of motion,
\begin{eqnarray}
\dot{\sigma}&=& \Gamma_R -(\Gamma_L+\Gamma_R) \sigma + \kappa x
+\xi_\sigma, \label{eq:eomP}
\end{eqnarray}
where $\Gamma_L$ and $\Gamma_R$ are the tunnel rates across the left
and right junctions, respectively, which can be varied by altering
the gate voltage. To get the equations of motion in the form of Eq.
(\ref{eq:fullLV}), we shift to variables describing deviations from
the steady state, $\sigma'=\sigma-\av{\sigma}$, and $x'=x-\av{x}$.
The evolution matrix ${\mathbf B}=\Gamma_L+\Gamma_R=\Gamma_T$ and
the coupling and fluctuation vectors $\vect{\kappa}=\kappa,
\vect\xi_\sigma=\xi_\sigma$ are simply scalars. Equation
(\ref{eq:gplusw}) then gives a simple expression for the damping and
frequency shift, $
\delta\omega_0^2(\omega)+i\omega\gamma_\sigma(\omega)=-\omega_0^2x_s\kappa(\Gamma_T-i\omega)/(\Gamma_T^2+\omega^2)$
and we use Eq. (\ref{eq:defG}), or the evolution of the first and
second moments \cite{RM}, to obtain,
\begin{eqnarray}
G_{\sigma\sigma}=
\Gamma_R(1-2\av{\sigma})+\Gamma_T\av{\sigma}+\kappa(\av{x}-2\av{x\sigma})\label{eq:G_SET_0b}.
\end{eqnarray}
This leads to the position and charge noise spectra, plotted in Fig.
1(a). The spectra will be thermal and Fano-like as long as weak
coupling holds $\Gamma_{L,R} \gg \gamma_T(\omega_0)$.

The SET-resonator equations are closed and linear to start with
\cite{SETa,SSET1}. In contrast, the equations of motion for a cavity
coupled to a resonator \cite{MCCG} are non-linear. We can still
apply our formalism if we linearize these equations about their
steady state values $\av{x},\av{a}$. With shifted variables
$x'=x-\av{x},a'=a-\av{a}$ and cavity detuning and damping, $\Delta$,
$\gamma_a$,
\begin{eqnarray}
\dot{v'}&=&-\omega_0^2x' -\gamma_{e}v'-2g (\langle a^\dag\rangle  a'+ a'^\dag \langle a\rangle ) +\xi_{e}\label{eq:cavres2a}\\
\dot{a'}&=&i(\Delta+g\av{x}) a' - \frac{\gamma_a}{2} a' +ig\: x'
\av{a}+\xi_a .\label{eq:cavres2b}
\end{eqnarray}
is obtained, plus the complex conjugate equation for the cavity
field ${a'}^\dag$.

The steady state values simply act as parameters, so Eq.
(\ref{eq:cavres2a}) has the same form as Eq. (\ref{eq:eomxv}). The
variable $\sigma_n= \langle a^\dag\rangle a'+ \langle a\rangle
a'^\dag$ is a particular quadrature of the field (determined by the
driving laser amplitude $\alpha$), which affects the resonator with
a strength determined by $x_s=2g/\omega_0^2$. We can also rewrite
Eq. (\ref{eq:cavres2b}) and its complex conjugate in the form of Eq.
(\ref{eq:eomsigma}) for the cavity variables $\vect{\sigma}=\left(\begin{smallmatrix} a'\\
a'^\dag
\end{smallmatrix}\right)$.
${\mathbf
B}=\left(\begin{smallmatrix} B^{-}&0\\
0&B^{+}\end{smallmatrix}\right)$ determines their mean evolution,
where $B^{\pm}=\frac{\gamma_a}{2}\pm i(\Delta+g\av{x})$, and
 $\vect{\kappa}=ig\left(\begin{smallmatrix} \langle a\rangle \\
-\langle a^\dag\rangle
\end{smallmatrix}\right)$, with
$\vect{n}^T=(\langle a^\dag\rangle, \av{a})$.
Cavity fluctuations are described by the standard terms, $G_{a^\dag
a}=\gamma_a{N_a}$, $G_{a a^\dag}=\gamma_a({N_a+1})$ where ${N_a}$
gives the cavity temperature. We insert $\mathbf{B,
G_{\sigma\sigma}}, \vect{n}^T$ and $\vect{\kappa}$ into Eq.
(\ref{eq:S_Dn}) to obtain the spectra. Weak-coupling holds when
$\gamma_a\gg\gamma_T(\omega_0)$:
 the cavity is damped much more
rapidly than the resonator.

%



Figure 1 shows the spectra of $x$ and $\sigma_n$ for the SET
resonator and the cavity resonator. As a comparison, we also plot
the spectra of a back-action free detector measuring a thermal
resonator with modified parameters \cite{SETa}. To calculate this we
eliminate the back-action term $\omega_0^2x_s\sigma_n$ in Eq.
(\ref{eq:eomxv}), calculate the spectra, and then replace the
thermal parameters with the values they would have due to
back-action (we set $\omega_0=\omega_r,\gamma_{e}=\gamma_{T},
D_e=D_T$ ). The resonator spectrum is very well matched, deviation
only occurring far from resonance where the noise is essentially
zero. In contrast, the thermal backaction model completely fails to
capture the anti-resonance in the device spectrum. A similar
antiresonance will occur for any such "detector" weakly coupled to a
resonator, if $\gamma_e$ and $D_e$ are low enough.
%

%

The frequency spectrum of the detector is a major way of obtaining
information about the dynamics of the resonator.  Although the
antiresonance occurs in one particular detector variable $\sigma_n$
(which might not be the most experimentally relevant), it should
 be possible to see signatures in other observables, e.g. a
dip in SET current (not charge) noise \cite{SETa,koerting}.

Input-output theory \cite{WM} $b_{out}=b_{in}+\sqrt{\gamma_a} a'/2$
shows the spectral properties of the cavity are directly transferred
to the output field, and hence detectable via a homodyne measurement
of a field quadrature. If $R\gtrsim1$, the output spectrum is well
approximated by the resonator spectrum multiplied by a
constant\cite{MCCG}. If the back-action noise is not negligible, it
will be modified to a Fano-like form. The spectrum will only take
the exact form Eq. (\ref{eq:S_Dn}) when the quadrature has the same
phase as $\sigma_n$, but the spectra of other quadratures will show
signatures of the Fano resonance, e.g. an asymmetric peak (Fig. 1).
%
%

A detector spectrum showing a Fano-lineshape can reveal more about
the dynamics of the resonator; the bare as well as renormalized
frequencies and both the total and environmental damping. Fano
resonances could lead to more sensitive measurements
 and applications such as transmission-line
switches and analogues of EIT \cite{vyurkov}.




In conclusion, we studied the spectra of a detector and resonator
coupled in a particular (but quite general) way. The detector
variables depend linearly on $x$, and the back action force is
proportional to a particular detector variable $\sigma_n$. If the
system can be described by linear Langevin equations, the spectra of
the resonator and the detector variable $\sigma_n$ can  easily be
derived. In the weak-coupling regime where the dissipation of the
detector is much larger than the total resonator damping (e.g.
$\gamma_a\gg\gamma_T$), these spectra reduce to a generic form.
 The
resonator spectrum is approximately thermal, but the spectrum of the
detector variable coupled to the resonator has a resonance at the
renormalized frequency and an anti-resonance at the bare resonator
frequency. We calculated the spectra of a resonator coupled
capacitively to a normal state SET, and parametrically to a
resonator. The detector spectra differ significantly from those
derived from a purely thermal back-action model.

We acknowledge helpful discussions with Andrew Armour, Steve Bennet,
Miles Blencowe, Aashish Clerk and Alex Rimberg, and funding by EPSRC
grant EP/D066417/1.


\begin{thebibliography}{99}
\bibitem{foot_1} If the master equations of the sub-systems are
simply coupled by a Hamiltonian term, Eq. (\ref{eq:gplusw}) equals
the anti-symmetrized noise, but is also valid if the dissipative
terms are modified (e.g. $x$-dependent tunnel rates).
\bibitem{foot_2} Eq. (\ref{eq:Snn}) has resonances at $\pm \omega_r$, not only
$\omega_r$ \cite{fano}, as we have not made the RWA, valid for a
slowly relaxing detector.
\end{thebibliography}

\end{document}